# Matrix representation of a solution of a combinatorial problem of the group theory

**Krasimir Yordzhev, Lilyana Totina**
*Faculty of Mathematics and Natural Sciences*

*South-West University*

*66 Ivan Mihailov Str, 2700 Blagoevgrad, Bulgaria*

E-mail: yordzhev@swu.bg , liliana_totina@abv.bg

**Abstract**: An equivalence relation in the symmetric group $S_n$, where $n \geq 2$ is a positive integer has been considered. An algorithm for calculation of the number of the equivalence classes by this relation for arbitrary integer $n$ has been described.

**Keywords:** Symmetric Group, Equivalence Relation, Equivalence Classes, Directed Graph, Euler Function, C++ Programming Language

## 1. INTRODUCTION

The object of the present study is to be received an algorithm for computer calculation of some combinatorial characteristics of the symmetric group. This algorithm has been based on the theoretical elaborations described closely in [3] and [4].

In our study with $\mathbb{N}$ we denote the set of the positive integers. Let $n \in \mathbb{N}$. Then with $\mathbb{Z}_n$ we denote the set
$$\mathbb{Z}_n = \{1, 2, \ldots, n\}.$$

With
$$S_n = \left\{ \begin{pmatrix} 1 & 2 & \cdots & n \\ i_1 & i_2 & \cdots & i_n \end{pmatrix} \middle| i_s \in \mathbb{Z}_n, s = 1, 2, \ldots, n, i_s \neq i_t \text{ for } s \neq t \right\}$$
we denote the symmetric group on the set $\mathbb{Z}_n$, i. e. the group of all one-to-one mappings $\alpha : \mathbb{Z}_n \to \mathbb{Z}_n$ of the set $\mathbb{Z}_n$ in oneself.

If $k, m \in \mathbb{N}$ then with $GCD(k, m)$ we denote the greatest common divisor of integers $k$ and $m$. Then $k$ and $m$ are relatively primes if and only if when $GCD(k, m) = 1$.

With $\varphi(n)$ we denote the Euler function, i. e. the number of elements of $\mathbb{Z}_n$ that are relatively primes with $n$ (see more for example in [1] or [5]). By definition $\varphi(1)=1$.

## 2. PRIOR INFORMATION

Let $n$ be a positive integer. We consider an element

$$\sigma = \begin{pmatrix} 1 & 2 & \ldots & n-1 & n \\ 2 & 3 & \ldots & n & 1 \end{pmatrix} \in S_n.$$

In [3] the following equivalence relation had been introduced in $S_n$: we say that $\alpha, \beta \in S_n$ are $\sigma$ - equivalent if the integers $k$ and $l$ exist such that $\alpha = \sigma^k \beta \sigma^l$, that is equivalent to condition that integers $k_1$ and $l_1$ exist such that $\sigma^{k_1} \alpha = \beta \sigma^{l_1}$. The following task has been put: To find the number of the equivalence classes by so defined equivalence relation, with another words the cardinality of the factor set $Q_n = S_{n/\sigma}$.

To be solved so putting task a directed graph $\Gamma_n$ with the set of vertices

$$V_n \subset \mathbb{Z}_n^2 = \{\langle a,b \rangle \mid a,b \in \mathbb{Z}_n\},$$

has been constructed, formed by the next mean:
1. For every integer $m \in \mathbb{Z}_n$ such that $GSD(n,m)=1$ a vertex $\langle 1,m \rangle \in V_n$ exists, as in this vertex arcs do not enter.
2. If $\langle k,l \rangle \in V_n$, then $k$ divides $n$.
3. Let $\langle k,l \rangle \in V_n$ and let $p$ is a prime divisor of $\dfrac{n}{k}$. Then we receive the number $r$ that is equal to the remainder after the multiplication $pl$ is divided into $n$. We construct a vertex $\langle pk,r \rangle \in V_n$ and an arc with begin the vertex $\langle k,l \rangle$ and end the vertex $\langle pk,r \rangle$.
4. Another vertex and another arc in the graph $\Gamma_n$ accepting received by the mean described above do not exist.

For example at $n=12$ the graph $\Gamma_{12}$ is shown on figure 1:

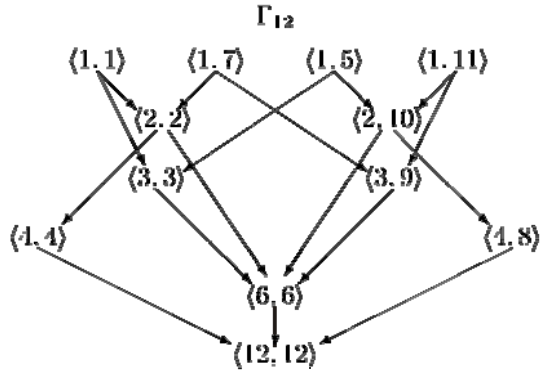

Fig. 1: The graph $\Gamma_{12}$.

We introduce the following partial order in the vertices set $V_n$ of the graph $\Gamma_n$: If $v_1, v_2 \in V_n$, then $v_1 < v_2$ if and only if when a directed path with begin vertex $v_1$ and end vertex $v_2$ exists. It is easily to see that by so introduced order $V_n$ is semilattice with the unique maximal element $\langle n,n \rangle \in V_n$ and $\varphi(n)$ in number minimal elements, each of kind $\langle 1,m \rangle \in V_n$, where $\varphi(n)$ is the Euler function.

For more details of the graph theory see for example in [6].

For each vertex $\langle k,l \rangle \in V_n$ we define the function $h(n,k)$ that depends on $n$ and $k$, but does not depend on $l$.

(1) $$h(n,k) = \frac{1}{k}\left((k-1)!\left(\frac{n}{k}\right)^{k-1} - \sum_{\langle r,t \rangle < \langle k,l \rangle} r\, h(n,r)\right)$$

The following assertion had been demonstrated in [3]:

**Theorem 1** [3] *The number of the equivalence classes by the $\sigma$-equivalence is equal to*

(2) $$|Q_n| = \sum_{\langle k,l \rangle \in V_n} h(n,k)$$

Theorem 1 gives us an effective algorithm for the manual calculation of $|Q_n|$. Construction of the graph $\Gamma_n$ is necessary for this object. This approach gives relatively good results at relatively small values of $n$ as the experience of authors has been shown. With increase of

$n$ probability of errors increases repeatedly, because of the number of classes increases exponentially, according to formulas (1) and (2).

It had been demonstrate in [4] that all assertions considered in [3] are valid for an arbitrary element $\sigma \in S_n$, on condition that $\sigma$ is a cycle with length $n$.

It is easy to see that we can reorganize the formula (1) in the following recursive kind:

$$(3) \quad h(n,k) = \frac{1}{k}\left[(k-1)!\left(\frac{n}{k}\right)^{k-1} - \sum_{\substack{r|k \\ r<k}} r\tau(n,k,r)h(n,r)\right] \text{ for } k>1$$

and

$$(4) \qquad h(n,1) = 1,$$

where the function $\tau(n,k,r)$ gives as the number of vertices $\langle r,s \rangle \in V_n$, such that $\langle r,s \rangle < \langle k,k \rangle \in V_n$.

The following assertion had been demonstrated in [4]:

**Theorem 2** [4] *The number of the vertices of kind $\langle k,l \rangle \in V_n$ is equal to $\varphi\left(\frac{n}{k}\right)$, where $\varphi(m)$ is the Euler function.*

As a consequence of theorem 2 the formula (2) has been reorganized in the following kind:

$$(5) \qquad |Q_n| = \sum_{k|n} h(n,k)\varphi\left(\frac{n}{k}\right)$$

## 3. MATRIX REPRESENTATION AND PROGRAM REALIZATION

In this section we use program language C++ for description of algorithms considered in the present study. The program was been tested in a programming environment Borland C++ Builder.

At the examples we consider that $n$ is a global constant integer parameter, equal to the order of the group $S_n$ and may be used without explicit declaration.

For using theorem 1 as we avail ourselves of formulas (3) and (4) it is convenient to fill the elements of the matrix $M = (m_{ij})$, consisting of 4 rows and $q$ columns, where $q$ is equal to the number of all positive divisors of the integer $n$ (including 1 and $n$). Never the less the integer will be written down in $M$, in our program realization we declare $M$ as two-dimensional array of type double because of fact that we expect to receive too large values at calculation of the function $h(n,k)$ according to formula (3).

In the first row we write down all integers dividing without remainder the parameter $n$. That can be realized with the help of the following procedure, for example:

```
void Divisors(int n, int FirstRow[], int& q)
{
 q = 1;
 FirstRow[0] = 1;
 for (int t = 2 ; t <= n; t++) {
    if (n%t == 0) {
       FirstRow[q] = t;
       q++;
    }
 }
}
```

As additional effect of the work of the function Divisors we receive the number $q$ of all divisors of the global parameter $n$. It the end the procedure receives an integer array FirstRow, correctly filled with the divisors of $n$. We copy the values of the array FirstRow to first row of the array $M$.

We fill the second row of $M$ with the values of the Euler function $\varphi\left(\frac{n}{k}\right)$, where $k$ has been taken from the corresponding component of the first row of $M$. The function $\varphi(m)$ can be realized as we use a variety of the algorithm, known as "the Sieve of Eratosthenes" [2,5]

```
int EulerFunction(int m) {
   int t =1;
   int b[100];
   for (int i=1; i <= m; i++) {
      b[i] = i;
```

```
        }
        for (int i=2; i <= m; i++) {
            if (b[i] == 0) {
                continue;
            }
            if (m%b[i] == 0) {
                for (int j=1; i*j<=m; j++) {
                    b[i*j] = 0;
                }
            }
            else t++;
        }
        return t;
    }
```

We fill the third row of $M$ with the values of the function $h(n,k)$, according to formulas (3) and (4), where $k$ has been taken from the corresponding component of the first row of $M$. Here we consider again that $n$ is a global constant parameter, i.e. it is not necessary $n$ to be present in the list of formal parameters of the function described below. For simplicity of the exposition we will not verify correctness of the data, i.e. whether $k$ divides $n$. This condition has been ensured by the fact that $k$ has been taken from the first row of the matrix $M$, where on condition only divisors of $n$ has been written down. We declare the type of the function h(int), as well as some working variables, which values are integers with the type long double because of expectation to receive to large values according to formula (3).

For the calculation of the function $h(n,k)$, must first be calculated the function $\tau(n,k,r)$. This can be done using the following algorithm:

```
    // Euclidean algorithm for finding the greatest common divisor of integers a and b:
    int GCD(int a, int b) {
     while (a != b)
        if (a > b) {
            a = a-b;
        }
        else b = b-a;
     return a;
    }

    //-------------------------------------------------------------------
```

// Function examined whether $\langle k,l \rangle$ is the vertex of $V_n$

```
bool Vertex(int k, int l, int n) {
  if ((k<=0) || (k>n) || (l>n) || (n%k != 0)) {
    return false;
  }
  else {
    if ((k == n)) {
      if (l == n) return true;
    }
    else
      for (int m=1; m<=n; m++) {
        if ((GCD(m,n) == 1) && (m*k%n == l)) return true;
      }
  }
  return false;
}
```

//----------------------------------------------------------------------

// The function $\tau(n,k,r)$:

```
int tau(int n, int k, int r) {
  int t=0;
  int p = k/r;
  for (int s=1; s<n; s++) {
    if (Vertex(r,s,n) && (s*p%n == k%n)) {
      t++;
    }
  }
  return t;
}
```

Then an algorithm for obtaining $h(n,k)$ according to formulas (3) and (4) can be realized using the following C++ function:

```
long double h(int n, int k) {
  long double t;
  if (k == 1) {
    t=1;
  }
```

```
    else {
       long double fact = 1;
       for (int i=2; i <= k-1; i++) {
          fact = fact*i;
       }
       long double pow = 1;
       int p = n/k;
       for (int i=1; i<=k-1; i++) {
          pow = pow*p ;
       }
       long double sum = 0;
       for (int r=1; r<=k-1; r++) {
          if (k%r == 0) {
             sum = sum+r*tau(n,k,r)*h(n,r);
          }
       }
       t = (fact*pow-sum)/k;
    }
    return t;
 }
```

We receive the fourth row of $M$ as multiply component by component the second with the third rom.

Here we will skip the description of the main function and the input and output operations as they are specific to different programming environments.

At $n=12$ the filled matrix $M$ is shown in table 1:

Tab. 1: Matrix $M$ at $n=12$.

| $k\mid n$ | 1 | 2 | 3 | 4 | 6 | 12 |
|---|---|---|---|---|---|---|
| $\varphi(n/k)$ | 4 | 2 | 2 | 2 | 1 | 1 |
| $h(n,k)$ | 1 | 2 | 10 | 39 | 628 | 3326054 |
| $m_{2j} * m_{3j}$ | 4 | 4 | 20 | 78 | 628 | 3326054 |

We receive the final result for the number $|Q_n|$ of the equivalence classes of $S_n$ by the considered equivalence relation as we add the elements of the last row of the matrix $M$ according to formula (5)

We calculate from table 1:
$$|Q_{12}| = 4+4+20+78+628+3326054 = 3326788$$

We show in table 2 the values of $|Q_n|$ at $n \leq 19$ calculated by the algorithm described above:

Tab. 1: The number of the equivalence classes in $S_n$ concerning to equivalence at $n \leq 19$.

| $n$ | 2 | 3 | 4 | 5 | 6 | 7 | 8 | 9 | 10 |
|---|---|---|---|---|---|---|---|---|---|
| $|Q_n|$ | 1 | 2 | 3 | 8 | 24 | 108 | 640 | 4492 | 36336 |

| $n$ | 11 | 12 | 13 | 14 | 15 |
|---|---|---|---|---|---|
| $|Q_n|$ | 329900 | 3326788 | 36846288 | 444790512 | 5811886656 |

| $n$ | 16 | 17 | 18 | 19 |
|---|---|---|---|---|
| $|Q_n|$ | 81729688428 | 1230752346368 | 19760413251956 | 336967037143596 |

**REFERENCES**

[1] Andreescu, T., Andrica, D., Feng, Z. (2007) 104 Number Theory Problems. Boston: Birkhauser.
[2] Reingold E. W. (1977) Combinatorial Algorithms. Theory and Practice. New Jersey, Prentice-Hall.
[3] Yordzhev, K. (2004) On an equivalence relation in the set of the permutation matrices, in Discrete Mathematics and Applications, Blagoevgrad, SWU "N. Rilski", 77-87.
[4] Yordzhev, K. (to appear) On a Combinatorial Problem in the Symmetric Group.
[5] Мирчев, И., (1995) Теория на числата, Благоевград, ЮЗУ „Н. Рилски".
[6] Мирчев И., (2001) Графи. Оптимизационни алгоритми в мрежи, Благоевград, ЮЗУ „Н. Рилски".